\documentclass[doublecol]{epl2}

\usepackage{amssymb, amsmath}
\begin{document}

\title{Delay before synchronization \& its role in latency of sensory awareness}

\author{Ad\`{e}le Peel\inst{1,2} \and Henrik Jeldtoft Jensen\inst{1, 2}}
\shortauthor{Ad\`{e}le Peel and Henrik Jeldtoft Jensen}

\institute{
  \inst{1} Institute for Mathematical Sciences, Imperial College London - 53, Princes Gate, London, UK, SW7 2PG\\
  \inst{2}Department of Mathematics, Imperial College London - 180 Queens Gate, London, UK, SW7 2AZ
}


\date{\today}

\abstract{
Here we show that for coupled-map systems, the length of the transient prior to synchronization is both dependant on the coupling strength and dynamics of connections:  systems with fixed connections and with no self-coupling display quasi-instantaneous synchronization. Too strong tendency for synchronization  would in terms of brain dynamics be expected to be a pathological case.  We relate how the time to  synchrony depends on coupling strength and connection dynamics to the latency between neuronal stimulation and conscious awareness. We suggest that this latency can be identified with the delay before a threshold level of synchrony is achieved between distinct regions within the brain, as suggested by recent empirical evidence, in which case the latency can easily be understood as the inevitable delay before such synchrony builds-up. This is demonstrated here through the study of simplistic coupled-map models.  
}
\maketitle
\section{\label{Introduction}Introduction}
When the brain receives a sensory stimulus, there is a delay before we become consciously aware of it \cite{Libet1964}\cite{Pockett2002}\cite{Gomes2002}. An explanation as to what causes the delay or what mechanisms occur during this latency is still outstanding. Separately, the moment of conscious awareness has empirically been associated with higher levels of synchrony between distinct brain regions \cite{Rodriguez1999}\cite{Melloni2007}. Here we show that for certain ranges of interaction strength, generic coupled systems display a delay between the time they begin to interact and when they achieve synchrony. We propose that the need for such a delay in these simple systems, and its dependence on the strength of interaction, can provide insight into the mechanisms at play in the brain during the latency between sensory stimuli and conscious awareness. If this latency is identified as the delay before a threshold level of synchrony is achieved between distinct regions within the brain, it can easily be understood since a delay is inevitable for such synchrony to build-up as demonstrated here by our simplistic models.

Current views of cognition, assume it to be a direct result of the physical \& chemical interactions of the millions of neurons that make up our brains. This notion has been extended in the Electromagnetic Field Theory of Mind, the essence of which can be understood as ``...consciousness is identical with certain spatiotemporal patterns in the electromagnetic field''\cite{PockettBook2000}.

Healthy cognition requires a careful balance between large-scale integration and distributed dynamics of anatomically and functionally segregated brain regions \cite{Varela2001}. Too much neuronal synchronization having been associated with known pathologies such as epilepsy \cite{EpilipsyBook1975}\cite{Mormann2000} and pathological tremors \cite{Freund1983}\cite{TremorBook1990}. Whilst too little synchronization has been linked to schizophrenia \cite{Lewis2005}. Other empirical evidence suggests that synchronization plays an important role in cognition \cite{Roelfsema1997}, in particular, synchronization plays an integral role in conscious awareness \cite{Rodriguez1999}\cite{Melloni2007}. So, for conscious awareness, a careful balance of the correct level of synchrony must be achieved: too little or too much could be pathological, yet enough is needed to raise awareness from sub-conscious to conscious. However, this has yet to add any insights to the contentious issue of the precise timing of conscious awareness.

The question of exactly when we become consciously aware of sensory stimuli is of course an ancient philosophical one, but has opened up to scientific investigation (and discussion) since the seminal work of Libet and co-workers (see \cite{MindTime} and references therein for details of the original research). Much attention has been given to the sparse data, with debate over Libet's conclusions lasting over many decades (see for example the special issue: \emph{Consciousness and Cognition, Vol 11, Issue 2 (2002)} and references therein).

Whilst some of Libet's conclusions still remain highly contentious (a good overview of the criticisms is given in \cite{Gomes2002} \& \cite{Gomes1998}), there is agreement, based on Libet's results and other backward masking experiments, that there exists a latency between sensory stimulation and conscious awareness of it \cite{Libet1964}\cite{Pockett2002}\cite{Gomes2002}. The precise duration of this latency is not confirmed, with estimates ranging from tens \cite{Pockett2002} to hundreds \cite{Libet1964} of milliseconds. It should, however, be noted that whatever the precise duration for any particular stimulus, in general, it is variable; dependant on a number of factors including the nature of the stimulus, whether it is direct cortical stimulation, visual \textit{etc}. and the intensity of the stimulus. Longer latencies being associated with lower intensity stimuli \cite{Libet1964}\cite{Gomes2002}.

Whilst the models used here are not intended to be interpreted as one-to-one models of neurons or even collections of neurons, it is expected that the same basic mechanisms in action within these simplified systems are also at play in more complicated and realistic neuronal models. By using simplified models we can gain a useful insight into these mechanisms that can be easily lost in the details of more realistic neuronal models, that would inevitably be more complicated. We show that basic coupled systems with co-evolving connections never synchronize instantaneously; they require a certain period of interaction for the synchronicity to build up. This period is longer for weaker interactions, similar to the latency between brain stimuli and conscious awareness being longer for weaker stimuli. If for conscious awareness to occur, a certain level of synchrony is required in the brain, we should in fact expect there to be a latency between stimulus and conscious awareness.
\section{\label{section: The Model and Results} The Model and Results}
\begin{figure}[!htbp]
\centering
\includegraphics[width=8.5cm,angle=0]{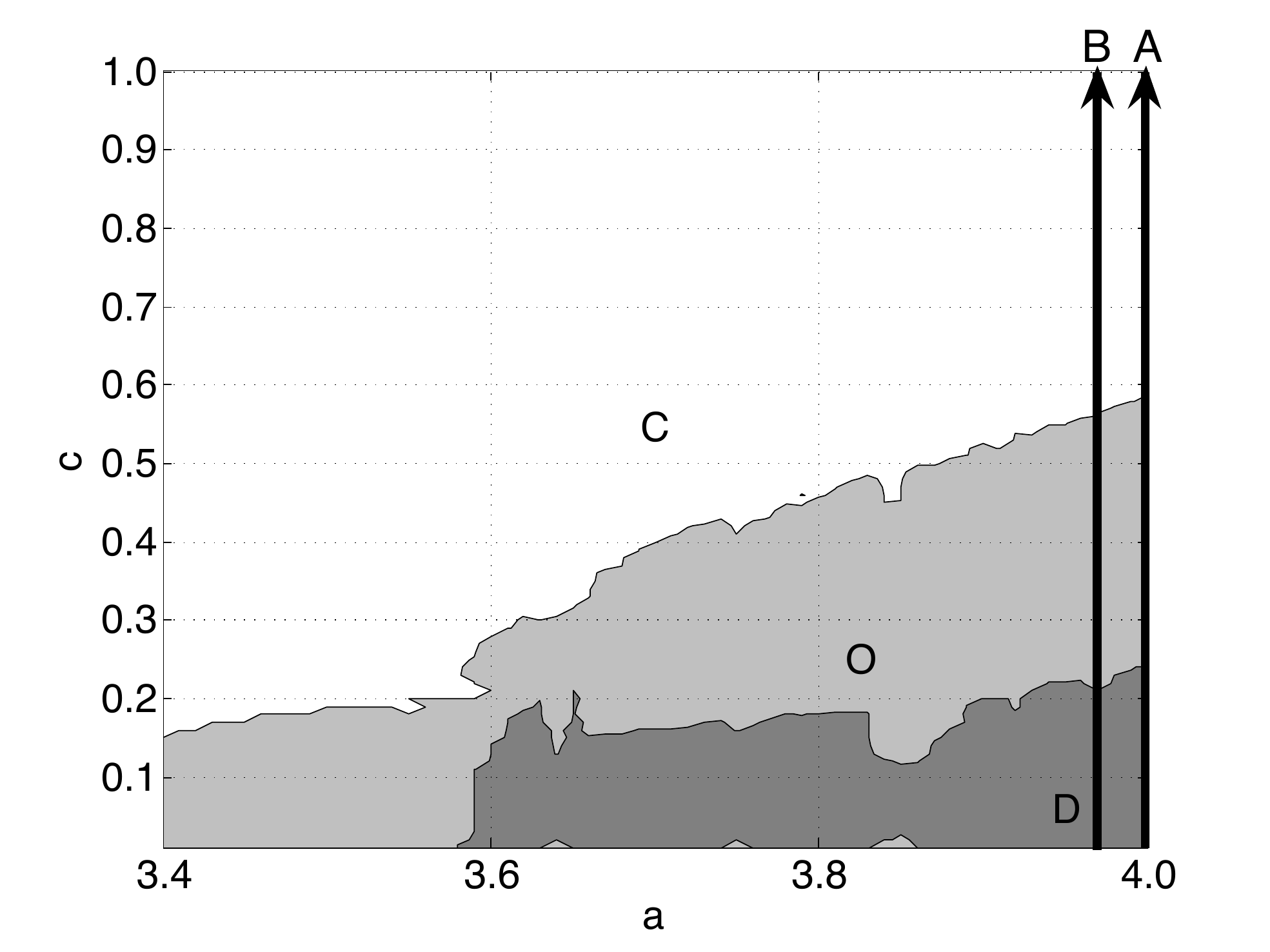}
\caption{Here is shown how the system of Eq.~(\ref{Equation: internally coupled Ito Kaneko model}) displays different global behaviours in different areas of the parameter space: coherent (C), the whole system synchronizes. Ordered (O), the system consists of several clusters within which the nodes are synchronized. Disordered (D), there is no synchronization between any pair of nodes, so their node-states all evolve independently. This is a reproduction of that shown as fig. 1 in \cite{ItoKaneko2003}.}\label{Figure: global parameter space}
\end{figure}

We compare the behaviour of two coupled-map systems: the first is a simple globally-coupled-map system with all-to-all connections. It consists of $N$ nodes that are each coupled to every other node with strength $\frac{c}{(N-1)}$. Each node's state, $x_n^i$, evolves according to:
\begin{equation}\label{Equation: internally coupled all-to-all fixed coupling}
x_{n+1}^i = f\Big[ (1-c)\cdot x_n^i + \frac{c}{N-1}\cdot \sum_{j=1}^N x_n^j \Big]
\end{equation}
\noindent The second is a more involved model, the basic distinction being that the connections between nodes are directional, weighted and co-evolve along with the nodes. They evolve according to Hebbian dynamics, so alike nodes have their connections strengthened at the expense of other connections between nodes that have states less alike. It is described by the following equations:
\begin{equation}\label{Equation: internally coupled Ito Kaneko model}
x_{n+1}^i = f\Big[ (1-c)\cdot x_n^i + c\cdot \sum_{j=1}^N w_n^{ij}\cdot x_n^j \Big]
\end{equation}
\noindent where the $w_n^{ij}$ evolve according to
\begin{equation}\label{Equation: link evolution}
w_{n+1}^{ij} = \frac{[1+\delta \cdot
g(x_n^i,x_n^j)]\cdot w_n^{ij}}{\sum_{j=1}^N[1+\delta \cdot
g(x_n^i,x_n^j)]\cdot w_n^{ij}}
\end{equation}
\begin{equation}\label{Equation: hebbian g}
g(x_n^i, x_n^j) = 1- 2\cdot |x_n^i - x_n^j|
\end{equation}
\noindent where $c\in [0,1]$ is the coupling strength, $x_n^i\in \mathbb{R}$, $i = 1,2,..., N$ and $w_n^{ij}$ denotes the weighted \& directional connectivity network. The parameter $\delta$ governs the plasticity of the network, we use $\delta=0.1$. We use the logistic map as the underlying map: $f(x)=ax\cdot(1-x$). This second model, Eq. \eqref{Equation: internally coupled Ito Kaneko model}, was previously studied in \cite{ItoKaneko2002}\cite{ItoKaneko2003}, where it was shown that the global parameter space splits into different phases, as shown in Fig.~\ref{Figure: global parameter space}.

\begin{figure}[!htbp]
\centering
\includegraphics[width=8.5cm,angle=0]{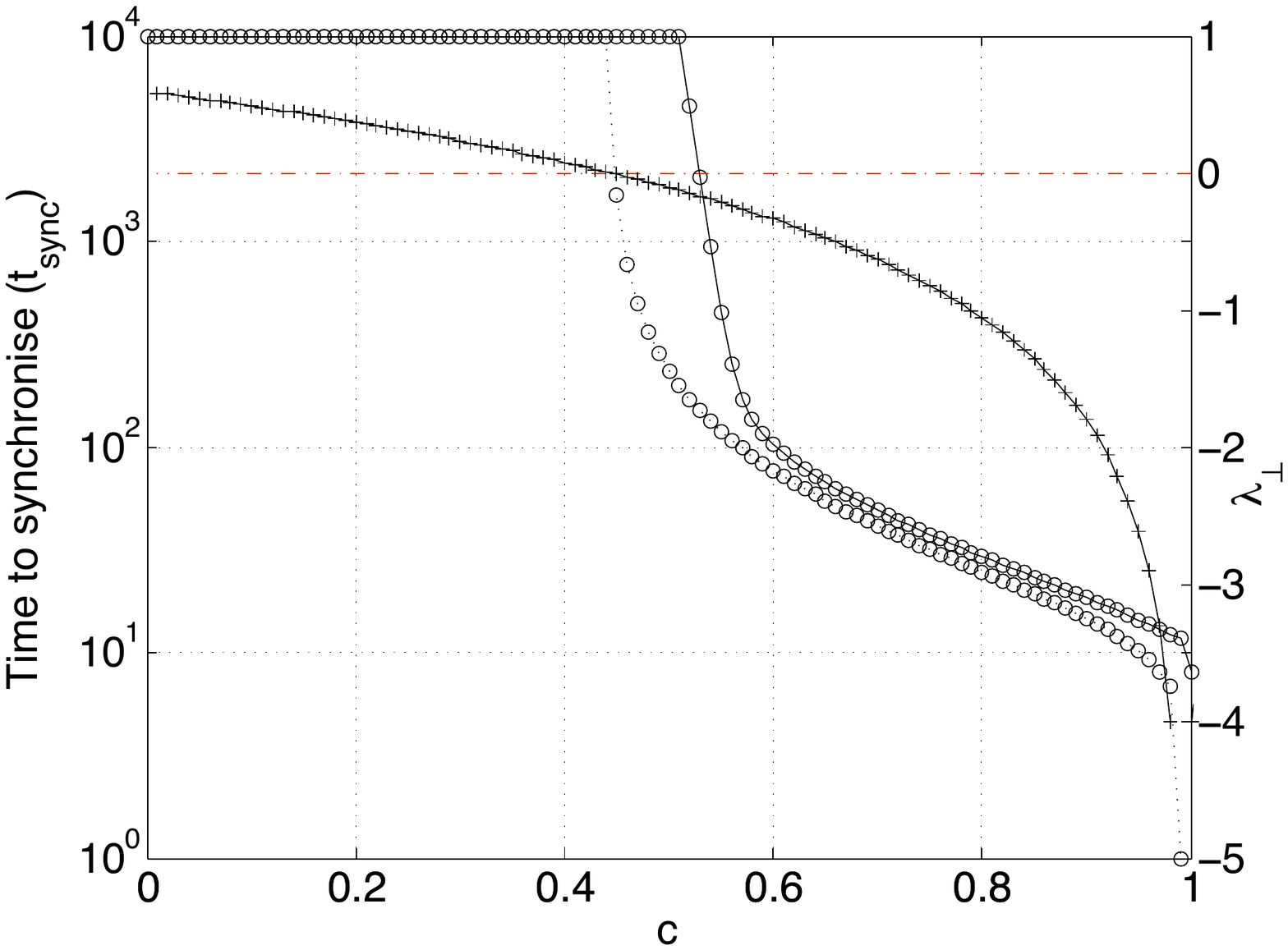}
\caption{The time it takes for the system with fixed couplings, see Eq. (\ref{Equation: internally coupled all-to-all fixed coupling}) (dashed line), and with dynamical couplings, see Eq. (\ref{Equation: internally coupled Ito Kaneko model}) (solid line), to re-synchronize after being perturbed. The data here is for different coupling strengths, $c$ and fixed nonlinearity parameter, \ensuremath{a=3.97}, as schematically indicated by arrow A of Fig. (\ref{Figure: global parameter space}). This particular $a$-value is that used in the previous studies \cite{ItoKaneko2002}\cite{ItoKaneko2003}. We show here the average time for re-synchronization of 100 random initial conditions. The system is perturbed each time the system synchronizes, with the synchronous threshold set as \ensuremath{\delta = 10^{-25}}. A value of \ensuremath{10^4} for the time-required-to-synchronize is indicative that no synchronization was recorded during the simulations. The solid line with  pluses indicates the Lyapunov exponent calculated from Eqs. (\ref{Lyapunov}) and (\ref{Lyapunov_0}).}\label{Figure: Time to sync a=3.97}
\end{figure}

\begin{figure}[!htbp]
\centering
\includegraphics[width=8.5cm,angle=0]{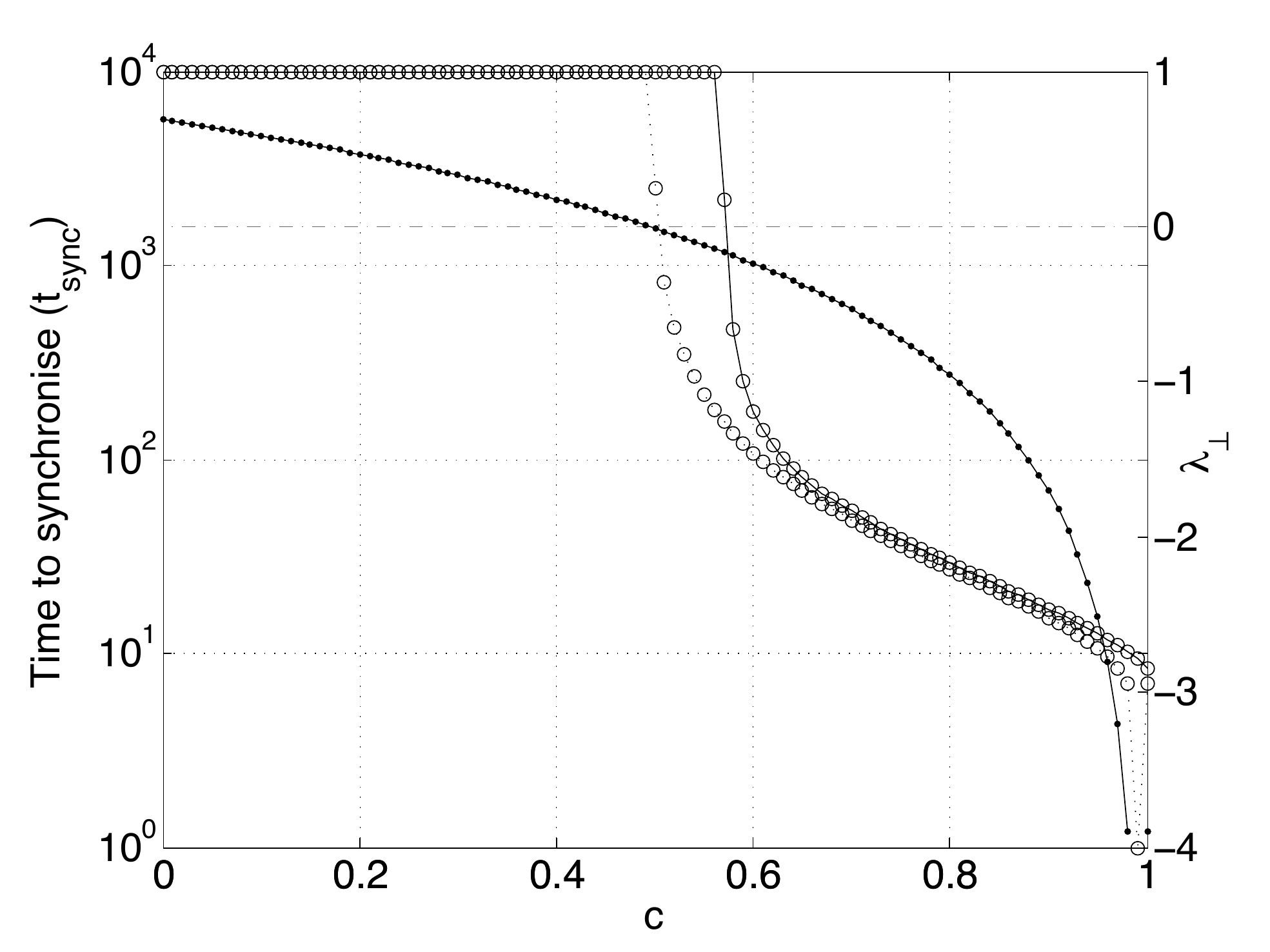}
\caption{The same as Figure~\ref{Figure: Time to sync a=3.97}, but for \ensuremath{a=4.00}, as schematically indicated by arrow B of Fig.~\ref{Figure: global parameter space}. This $a$-value is chosen as it is the only rational number of parameter value for which an absolutely continuous invariant probability measure has been proven to exist, ensuring a well defined Lyapunov exponent.The solid line with  bullets indicates the Lyapunov exponent calculated from Eqs. (\ref{Lyapunov}) and (\ref{Lyapunov_0})}\label{Figure: Time to sync a=4.00}
\end{figure}

\begin{figure}[!htbp]
\centering
\includegraphics[width=8.5cm,angle=0]{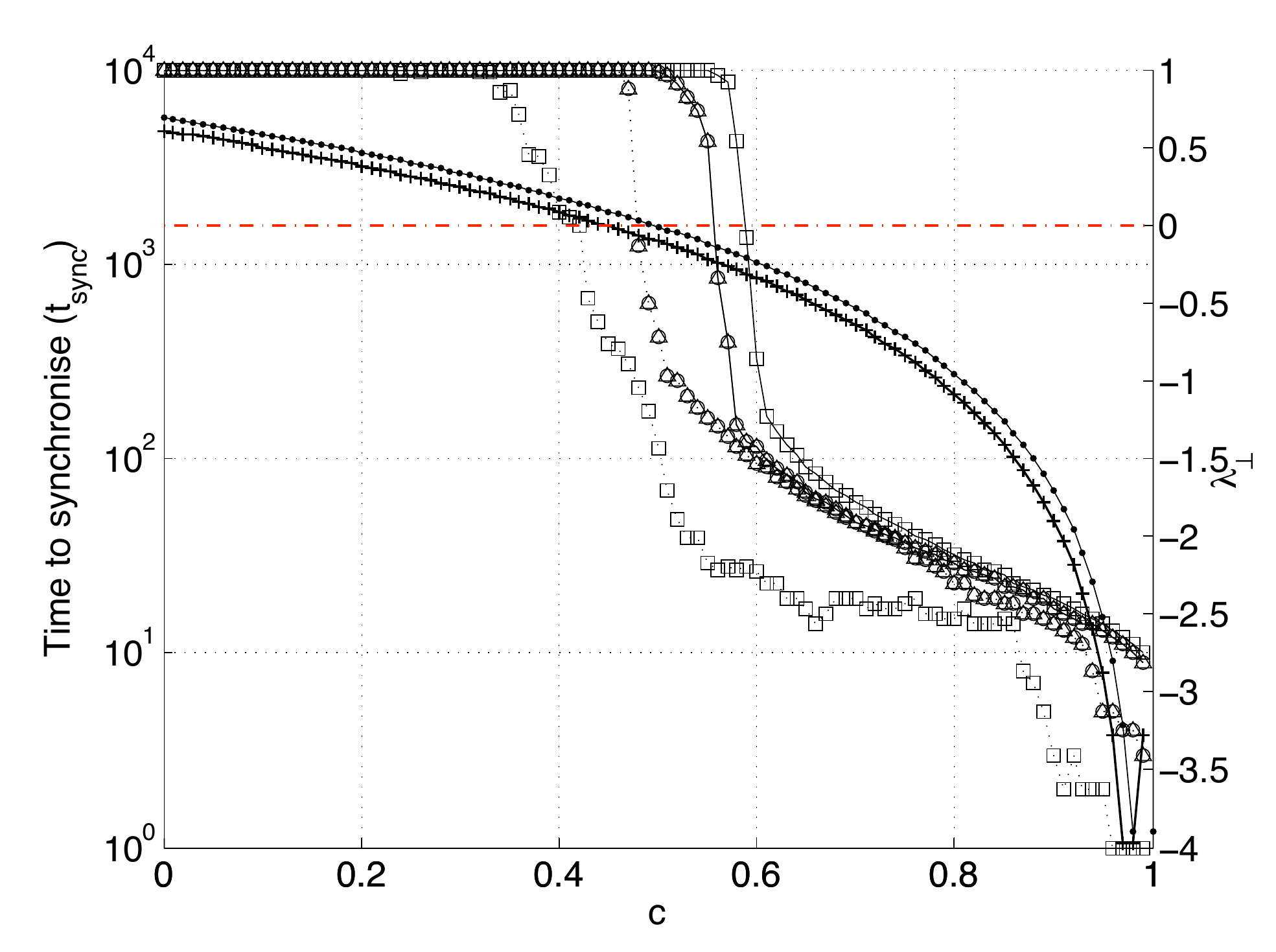}
\caption{The time it takes for the driven systems to synchronize, with the synchronous threshold, \ensuremath{\delta = 10^{-25}}. Each system consists of 100 nodes, 5 of which receive the external stimulus. The solid lines are for the system with co-evolving connections whilst the dashed lines are for the system with fixed all-to-all connections. The stimuli are as follows: (1) drive type A, with \ensuremath{\omega = 0.28} and \ensuremath{a = 3.97} (\ensuremath{\triangle}); (2) drive type A, with \ensuremath{\omega = 0.28} and \ensuremath{a = 4.00} (\ensuremath{\square}); (3) drive type B, with \ensuremath{a_{stimulus} = 4.0} and \ensuremath{a = 3.97} (\ensuremath{\bigcirc}). All data is shown as the average over 100 random initial conditions. A value of \ensuremath{10^4} for the time-required-to-synchronize is indicative that no synchronization was recorded during the simulations.The solid line  (pluses $a=3.97$, bullets $a=4.00$) indicates the Lyapunov exponent calculated from Eqs. (\ref{Lyapunov}) and (\ref{Lyapunov_0})} \label{Figure: time to sync for stimulated system}
\end{figure}

To relate the time to synchronization and the stability of the time evolution we also measure the transverse Lyapunov exponent according to(see Ref. \cite{ItoKaneko2003})
\begin{equation}
\lambda_\perp = \ln\Big| 1 - \frac{c\cdot N}{N-1} \Big| + \lambda_{0}
\label{Lyapunov}
\end{equation}
\noindent where $\lambda_0$ is the Lyapunov exponent of the underlying map $f$ calculated from the usual expression\cite{Strogatz}
\begin{equation}
\lambda_0=\lim_{T\rightarrow \infty} \frac{1}{T} \sum_{n=1}^T \ln \Big| f'(x_n^i) \Big|,
\label{Lyapunov_0}
\end{equation}
where we use $T=10^4$ and have checked this value to be sufficiently large for $a<4$. At $a=4$ we use the exact value $\lambda_\perp=\ln(2)$, see Ref. \cite{Finch}.
 For the logistic map, $\lambda_0$ is highly dependent on the nonlinearity parameter $a$, so in the figures 2, 3 and 4 below, we plot the values of $\lambda_0$ corresponding to the $a$-values used for the underlying map of the dynamical nodes in Eqs.~\eqref{Equation: internally coupled all-to-all fixed coupling} and \eqref{Equation: internally coupled Ito Kaneko model}. The transverse Lyapunov exponent denotes the stability of the synchronous state for the fixed system of Eq.~\eqref{Equation: internally coupled all-to-all fixed coupling}: positive (negative) values corresponding to the synchronous state being unstable (stable). It has previously been assumed that this same transverse Lyapunov exponent also denotes the stability of the synchronous state of Eq.~\eqref{Equation: internally coupled Ito Kaneko model} \cite{ItoKaneko2003}, since when the system displays coherent behaviour, the connections become static and almost homogeneous with value $w^{ij}_n\frac{1}{(N-1)}$. Below we show the Lyapunov exponent together with the time to synchronization. We would expect the coupled maps to be able to synchronize when $\lambda_\perp<0$. We find indeed that change of sign of the Lyapunov exponent obtained from Eqs. (\ref{Lyapunov}) and (\ref{Lyapunov_0})   coincides with the divergence of the time to synchrony  for the case with constant couplings, see the dashed line in  Figs.~\ref{Figure: Time to sync a=3.97}, \ref{Figure: Time to sync a=4.00} and \ref{Figure: time to sync for stimulated system}. However, the divergence of the time to synchrony for the system with dynamical couplings does not coincide with the change of sign of $\lambda_\perp$ derived from Eqs.  (\ref{Lyapunov}) and (\ref{Lyapunov_0}).  From this we conclude that the true Lyapunov exponent for the interacting system is not sufficiently accurately represented by  Eqs. (\ref{Lyapunov}) and (\ref{Lyapunov_0}). Unfortunately it is not possible to obtain an accurate numerical estimate from the Lyapunov exponent for the fully interacting system by direct averaging over the orbits. The computational processing required is prohibitive, since a reasonable density of points across the entire phase space is needed. In the coupled systems, it is nontrivial to get such a density of points in the 100 dimensional phase space.
  
The simpler system of Eq.~\eqref{Equation: internally coupled all-to-all fixed coupling} also displays different behaviours throughout the parameter space, the interested reader is referred to \cite{Kaneko1990} for further details. Here, we are concerned with the time it takes for the system to reach the Coherent state where all nodes are completely synchronized, so we shall not dwell further on the specifics of these models as the qualitative behaviour we focus on is believed to be a general feature of all coupled systems able to reach a state of synchronicity. In the case of systems defined in Eqs. \eqref{Equation: internally coupled all-to-all fixed coupling} \& \eqref{Equation: internally coupled Ito Kaneko model}, we define them to be synchronized if
\begin{equation}
\sigma^2 = \lim_{n\longrightarrow \infty} \sum_{\textmd{all pairs }i,j} |x_n^i-x_n^j| =0
\end{equation}
In practice, for numerical studies, we must consider systems to be synchronized if $\sigma^2$ is less than some threshold $\delta_{thr}$. The results shown in Figs.~\ref{Figure: Time to sync a=3.97} and \ref{Figure: Time to sync a=4.00} show how long it takes for the system to re-synchronize: each time the system reaches the synchronous state, each node-state, $x_n^i$, is perturbed by a random number $\eta<10^{-2}$, whilst ensuring that $x_n^i\in[0,1]$ is not violated.
 
From Figs.~\ref{Figure: Time to sync a=3.97} and \ref{Figure: Time to sync a=4.00} we see that the time required before synchrony is achieved varies dramatically for different coupling strengths. Neither system synchronizes (before the cut-off time of $10 ^4$) for low coupling strength, but is able to synchronize for coupling strengths greater than some critical coupling strength. For the values of coupling strength where the systems synchronize, the time it takes them to synchronize decreases as the coupling strength increases. It should also be noted that for all values of the coupling strength, $c$, the system with fixed all-to-all coupling of Eq.~(\ref{Equation: internally coupled all-to-all fixed coupling}), takes fewer iterations to synchronize than does the system with co-evolving connections. This becomes quite a stark difference when the system with fixed connections achieves synchrony quasi-instantaneously at $c=0.99$. By quasi-instantaneous we mean that due to the discrete nature of the system, it is not possible to record a shorter time difference than $1$ timestep. Indeed, for $c=0.99$, we observe synchronization from one timestep to the next.

Both systems are also able to synchronize when a random subset of the nodes are driven by an external stimulus. We imagine this mode of driving to be related to how dynamics of the brain responds to stimula. We implement the driving by choosing a random subset of nodes to receive the external drive, whose node-state evolution is altered as follows:
\begin{equation}\label{Equation: internally coupled Ito Kaneko model with drive}
x_{n+1}^i = f\Big[ (1-c)\cdot x_n^i + c\cdot \sum_{j=1}^N w_n^{ij}\cdot x_n^j + \frac{c}{N_{Tot.}-1}\cdot x_n^{drive} \Big]
\end{equation}
\noindent where $N_{tot.}$ is the sum of the number of nodes and the number of stimuli. Fig.~\ref{Figure: time to sync for stimulated system} shows the time required to achieve synchrony for different coupling strengths for these driven systems with two types of external stimuli:
\begin{equation*}
{\rm (A)}\;\; x_n^{drive} = \sin^2(\omega n)
\end{equation*}
\begin{equation*}
{\rm (B)}\;\; x_{n+1}^{drive} = a_{stimulus}\cdot x_n^{drive}\cdot(1-x_n^{drive})
\end{equation*}
\noindent The first drive type, A, is a simple oscillatory drive and the frequencies used are chosen as inverse of the average return time through a Poincar{\' e} section for the undriven  system. However we note that similar responses are seen for other frequencies. The second drive type, B, is a logistic map with mismatched nonlinearity parameter ($a_{stimulus}$). This is used as there has been evidence to suggest that functions with the form of the logistic map may indeed be of direct relevance in neuroscience \cite{Kuhn2004}.

Similar to the undriven system, no synchrony occurs for low coupling strength. Whereas, for higher coupling strengths, the systems are again able to synchronize for coupling strengths larger than a critical strength, $c_{crit.}$. For both drive types, A and B, the system with fixed connectivity takes shorter times to synchronize. When the fixed connectivity system is driven by drive type B, it again displays the quasi-instantaneous synchronicity for $c\gtrsim 0.96$.
\section{\label{Discussion}Discussion}
Whilst these simple models are not intended to be, nor could they ever be conceived to offer, a one-to-one model of neuronal dynamics, the general approach to synchronization is of relevance to the question of what mechanisms cause the delay between sensory stimuli and conscious awareness. From such simple systems, we hope to gain insight into the mechanisms at play in more complicated, physically realistic models. It is believed that this basic result of a delayed onset of synchronization, as displayed by these simple models, is a universal feature amongst coupled systems capable of synchronicity.

For real neural systems we expect the regime of $c$ close to one to be most relevant as this corresponds to a large number of in put from other neurones. With this in mind we note that the observation that the system with fixed coupling is able to achieve synchrony quasi-instantaneously is suggestive that time-evolving connections are of utmost importance within neuronal models. Only with co-evolving connections is it guaranteed that quasi-instantaneous synchronization does not occur. The behaviours of these models suggest however, that it is perhaps necessary to have a comparatively higher strength of connection in order to achieve the same duration of delay before synchronization if the connections co-evolve along with the nodes as compared with having fixed connectivity.
 
When looking to real neuronal connections, it is expected that they should not be static in time. Indeed, changes are observed in connectivity across many timescales. We have here highlighted the difference to timings of synchrony that such time-varying connections can make.

It is also interesting to note the potential extrapolations and insights possible from a greater understanding of the interplay between the levels of synchronization in the brains of epileptics and potential underlying causes. Could it for example be possible that epileptics have less time-dependence in their neuronal connectivity? Does this allow them to have faster reaction times as a consequence? Future work is required into the existence of similar latencies before synchrony in more neurologically accurate models, as well as other simple empirical observations in order to answer these questions with certainty.

It still remains, however, that coupled systems require a critical level of interaction, above which, the time it takes for them to synchronize decreases as the coupling strength is increased. This is of importance for understanding the latency between sensory stimuli and conscious awareness of this stimulus. Indeed, if conscious awareness is identified with gaining threshold levels of synchronization as implied by recent empirical evidence \cite{Melloni2007}, we should actually expect there to be a delay between the stimulus and conscious awareness, since in any system with time-varying connections, it takes time for synchronicity to build up.
\section{\label{Acknowledgements}Acknowledgements}
Ad\`{e}le Peel gratefully acknowledges the Engineering and Physical Sciences Research Council (EPSRC) for her Ph.D. studentship.

\end{document}